\RequirePackage{amsmath}
\documentclass{article}
\usepackage{kr}
\usepackage{times}
\usepackage{soul}
\usepackage{url}
\usepackage{hyperref}
\usepackage[utf8]{inputenc}
\usepackage{graphicx}
\usepackage{booktabs}
\usepackage{algorithm}
\usepackage{algorithmic}
\usepackage{listings}
\usepackage{todonotes}
\usepackage{enumitem}
\usepackage{tikz}
\usetikzlibrary{shapes, arrows, calc}

\tikzset{ 
	coggear/.pic={
		\def\teethnumber{8}     
        \def\threadheight{1.5mm}   
        \def\outrad{5mm} 
        \draw[fill=gray!50,even odd rule] 
        let 
        \n{dpt} = {360/\teethnumber)}
        in
        (0,0) circle (3mm)                          
        ({\n{dpt}*(0.5)}:\outrad) \foreach \x in {1,...,\teethnumber}{ 
        arc ({\n{dpt}*(\x-0.5)}:{\n{dpt}*(\x-0.25)}:\outrad)  
        --++(\x*\n{dpt}:\threadheight)                        
        arc ({\n{dpt}*(\x-0.25)}:{\n{dpt}*(\x+0.25)}:\outrad) 
        --++(\x*\n{dpt}:-\threadheight)                       
        arc ({\n{dpt}*(\x+0.25)}:{\n{dpt}*(\x+0.5)}:\outrad)--
        ({\n{dpt}*(\x+0.5)}:\outrad)
        };
	}
}

\definecolor{MyLightGray}{RGB}{230, 230,230}
\definecolor{MyDarkBlue}{RGB}{10, 10, 185}
\definecolor{MyCyan}{RGB}{20, 145, 145}
\definecolor{MyDarkGreen}{RGB}{0, 175, 60}
\definecolor{MyDarkRed}{RGB}{175, 0, 0}
\definecolor{MyBrown}{RGB}{165,42,42}
\newcommand{\CodeSymbolGreen}[1]{\textcolor{MyDarkGreen}{#1}}

\newcommand{\CodeSymbolCyan}[1]{\textcolor{MyCyan}{#1}}
\newcommand{\CodeSymbolRed}[1]{\textcolor{MyDarkRed}{#1}}
\newcommand{\CodeSymbolBrown}[1]{\textcolor{MyBrown}{#1}}

\lstdefinestyle{codestyle}{
    basicstyle=\ttfamily\scriptsize,
    breakatwhitespace=false,         
    breaklines=true,                 
    captionpos=b,                    
    keepspaces=true,                 
    numbers=left,                    
    tabsize=2
}

\lstdefinelanguage{rule}
{   
    numbersep=3pt, xleftmargin=0.3cm, xrightmargin=0cm,
    columns=fullflexible,
    keywordstyle=\color{red},
    keywords=[1]{@prefix,PREFIX,@base,BASE,SELECT,DISTINCT,ORDER,BY, VALUES, FILTER, WHERE,UNION, GROUP, BIND, INSERT, ASK, SERVICE},
    morekeywords=[2]{{[},{]}},
    morekeywords=[2]{{\{},{\}}},
    breaklines=true,
    comment=[l]{\#},
    morecomment=[s][\color{blue}]{<}{>},
    tabsize=4,
    alsoletter={-?}, 
    morestring=[b][\color{black}]\",
    showstringspaces=false,
     literate= {;}{{\CodeSymbolRed{;}}}1
     {.}{{\CodeSymbolRed{.}}}1
     {"}{{\CodeSymbolRed{"}}}1
     {\{}{{\CodeSymbolGreen{\{}}}1
     {\}}{{\CodeSymbolGreen{\}}}}1
     {]}{{\CodeSymbolCyan{]}}}1
     {[}{{\CodeSymbolCyan{[}}}1
     {(}{{\CodeSymbolBrown{(}}}1
     {)}{{\CodeSymbolBrown{)}}}1,  
    moredelim=[s][\color{MyDarkBlue}]{:}{\ },
    moredelim=[s][\color{MyDarkRed}]{@}{\ },
}

\tikzstyle{terminator} = [rectangle, draw, text centered, rounded corners, minimum height=1em, font=\footnotesize]
\tikzstyle{process} = [rectangle, draw, text centered, minimum height=1em, font=\footnotesize]
\tikzstyle{decision} = [diamond, draw, text centered, minimum height=1em, font=\footnotesize]
\tikzstyle{data}=[trapezium, draw, text centered, trapezium left angle=60, trapezium right angle=120, minimum height=1em, font=\footnotesize]
\tikzstyle{connector} = [draw, -latex']

\title{Eat your own KR: a KR-based approach to index Semantic Web Endpoints and Knowledge Graphs}

\author{%
Pierre Maillot$^1$\and
Catherine Faron$^1$\and
Fabien Gandon$^1$\and
Franck Michel$^1$\and
Pierre Monnin$^1$\\
\affiliations
$^1$Université Côte d'Azur, Inria, CNRS, I3S, Sophia-Antipolis, France\\
\emails
\{first.last\}@inria.fr
}

\begin{document}

\maketitle

\begin{abstract}
Over the last decade, knowledge graphs have multiplied, grown, and evolved on the World Wide Web, and the advent of new standards, vocabularies, and application domains has accelerated this trend.
%
IndeGx is a framework leveraging an extensible base of rules to index the content of KGs and the capacities of their SPARQL endpoints.
In this article, we show how knowledge representation (KR) and reasoning methods and techniques can be used in a reflexive manner to index and characterize existing knowledge graphs (KG) with respect to their usage of KR methods and techniques.
We extended IndeGx with a fully ontology-oriented modeling and processing approach to do so. 
Using SPARQL rules and an OWL RL ontology of the indexing domain, IndeGx can now build and reason over an index of the contents and characteristics of an open collection of public knowledge graphs.
Our extension of the framework relies on a declarative representation of procedural knowledge and collaborative environments (e.g., GitHub) to provide an agile, customizable, and expressive KR approach for building and maintaining such an index of knowledge graphs in the wild.
In doing so, we help anyone answer the question of what knowledge is out there in the world \textit{wild} Semantic Web in general, and  we also help our community monitor which KR research results are used in practice.
In particular, this article provides a snapshot of the state of the Semantic Web regarding supported standard languages, ontology usage, and diverse quality evaluations by applying this method to a collection of over 300 open knowledge graph endpoints.
\end{abstract}
\section{Introduction}

As of September 2023, the \href{https://lod-cloud.net/}{LOD cloud} was listing more than 1 300 knowledge graphs (KGs) published according to the Linked Data principles, with more than 16~300 links between them.
Such KGs consist of one or several datasets and can be accessed with one or several methods, including via SPARQL endpoints or as dumps.
Ontologies and KGs are thus scattered on the web and are accessible in various ways.
This growing number of KGs on the web presents an opportunity, as one of the first questions that arises when building a Knowledge Representation (KR) system is identifying existing resources on the web and their overlaps with the KR system being built, for possible reuse.
However, due to the scattering and diverse access means of KGs, existing indexing solutions are of limited help in answering this simple, initial question.

Therefore, in this paper, we wish to illustrate how the Knowledge Representation and Reasoning (KRR) technologies and methods can be used successfully to build an enhanced index of the Semantic Web itself. This index can be leveraged to answer some open questions regarding the available KGs, such as: 
What are their characteristics in terms of content, expressivity, quality, etc.?
What are the representation formalisms they use? 
What are the reasoning features that they support?

\begin{figure}[b]
    \begin{center}
        \begin{tikzpicture}[font=\footnotesize]

            \draw [rounded corners] (-0.5, 0.35) rectangle (4, 2.5) {};
            
            \draw [rounded corners, fill=green!10] (-0.25, 0.5) rectangle (0.5, 1.9) {};
            \draw [rounded corners, fill=green!10] (0.75, 0.5) rectangle (1.5, 1.9) {};
            \draw [rounded corners, fill=green!10] (1.75, 0.5) rectangle (2.5, 1.9) {};
            \node [rotate=90] at (0.125, 1.2) {Dataset 1};
            \node [rotate=90] at (1.125, 1.2) {Dataset 2};
            \node [rotate=90] at (2.125, 1.2) {Dataset 3};
            \node at (3.3, 1.25) {$\dots$};
            \node at (1.75, 2.25) {Knowledge graph }; 

            \draw [rounded corners, dotted] (5.25, 0.35) rectangle (7.25, 2.5) {};
            \node at (6.25, 2.25) {Endpoint};
            \pic[scale=0.8] at (6.25,1.25) {coggear};

            \draw[<->, >=latex, line width=1pt] (4.05,1.25) -- (5.2,1.25);
            \node at (4.625, 1.6) {access};
        \end{tikzpicture}
    \end{center}
    \caption{The couples ``knowledge graph -- endpoint'' found on the Web as considered in this paper.}
    \label{fig:kg_def}
\end{figure}
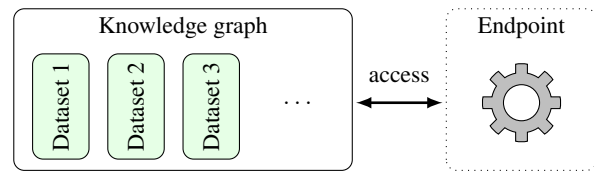

In the rest of this paper, we consider KGs published through a SPARQL endpoint (Figure~\ref{fig:kg_def}).
We propose to see an index of KGs as a regular KG of its own, and the problem of indexing KGs in the wild as an application domain for KR.
We leverage the models and methods from KRR to support each step of the indexing lifecycle (Figure~\ref{fig:indegx_indexation}): model and formalize (with ontologies, rules, and KGs), extract and populate KG descriptions, validate and augment the descriptions, infer and enrich the descriptions and, finally, publish the index.
Moreover, we set out to do so in the wild with real public KGs, dealing with real open endpoints, real-world data, volumes, heterogeneity in contents and implementations, etc.
This work is an extensive extension of the IndeGx framework presented in the journal article \cite{maillot2023indegx}, to which we added 
a model and method (OWL-RL ontology to describe endpoints), which results in 
an extended available IndeGx KG. In addition, this extension allows us 
to provide new results on KR usages in LOD datasets. 

\section{Related Works}

The indexing of KGs in the wild is a recurrent research subject, with multiple approaches depending on the domains, use cases, and requirements that are considered and, in particular, on the tasks supported by the resulting indexes (e.g., source selection, distributed query planning, distributed caching, and mirroring, etc.).

First and foremost, publishing KGs is generally done by hosting the data online, followed by submitting that source to some index services. Annotations are usually added to the KGs themselves (e.g., using VoID, DCAT, Creative Commons, or Dublin Core vocabularies) to feed such indexes, and these annotations are often generated manually. Notable examples of index services include the LOD cloud\footnote{\url{https://lod-cloud.net/}} for KGs in general or Linked Open Vocabularies\footnote{\url{https://lov.linkeddata.es/dataset/lov}}~\cite{vandenbussche2017LOV} for ontologies. The LOD cloud, in particular, is a prominent source of KGs for indexing approaches. However, it is only updated yearly, and its content is not directly accessible as a KG. Yet, most of the approaches to evaluate the KGs in the wild result from a one-shot campaign and often rely on these central indexes of KGs, giving us a snapshot of the wild Semantic Web at a given time.

Among such campaigns, several articles  proposed studies on the usage of vocabularies in datasets.
\cite{brinkmann2023webdatacommonsschema} proposes to extract a collection of statistics on the usage of the Schema.org vocabulary from RDF datasets embedded in Web pages. 
\cite{shi2023VOYAGELargeCollection} provides insight into the usage of vocabularies available through different data portals. 
Before them, \cite{glimm2012OWLArriveWeb} gave insight into the usage of OWL itself in the KGs of the Semantic Web.

Other initiatives focused on KGs specific characteristics. For example, GeoLOD~\cite{kopsachilis2020geolod} offers an online index of KGs from the LOD cloud annotated with their geographical coverage. At the same time, \cite{polleres2020MoreDecentralizedVision} surveys the actual availability of the KGs found in the LOD Cloud.
All the previously cited approaches either do not interact with KGs directly or download a static dump of the KGs' datasets to treat them locally.

An alternative is to interact with KGs through their SPARQL endpoints. This guarantees that the indexed KGs and their infrastructures are always in their latest version and supports the provision of fresh indicators, the constant monitoring of the KGs and their services, and the study of their evolution over time. Among such approaches, we find several monitoring-specific aspects such as  SPARQLES~\cite{vandenbussche2017sparqles} and YummyData~\cite{yamamoto2018yummydata} that evaluate the support of SPARQL by the endpoints and their general performances.

Our contribution in this paper is based on the IndeGx framework~\cite{maillot2023indegx}.
The IndeGx framework leverages KR and semantics to build, represent, and enrich an index of publicly available knowledge graphs on the Semantic Web. 

In \cite{maillot2023indegx},
the IndeGx framework is compared to existing approaches. 
It is shown that IndeGx replicates most of the evaluations used by other approaches. Additionally, using a standard format, the declarative nature of IndeGx rules guarantees its interoperability, extensibility, and versatility. Approaches such as SPARQLES encapsulate their logic in their code, hindering such characteristics natively offered by IndeGx.

The review of the approaches related to IndeGx in \cite{maillot2023indegx} shows that there is a growing interest in the extraction of metadata from KBs for various usages. IndeGx is a transparent and extensible declarative tool to extract metadata based on SPARQL queries, including all the metadata extracted by SPORTAL, SPARQLES, and LOUPE. It also provides original metadata, among which are several statistics on the usage of the namespaces in resource URIs and literal datatypes, the extraction of the timezone of the server, and a few SPARQL features not covered by SPARQLES.

The reproduction, the combination, and the extension of previous initiatives, showed the ability of IndeGx to smoothly integrate new features which makes it suitable for different uses: indexing, monitoring, and exploration. This ability to integrate new features and the transparency of its generation process are key features that distinguish IndeGx from other related approaches.

In this paper, we rely on public SPARQL endpoints to access the KGs and assess them in their latest version and state. In addition, we also use a KG to represent and specify the interaction with each endpoint, the features we want to extract, and the strategies to obtain them. Moreover, the resulting index is itself represented and published as a KG with a public endpoint for access -- the circle is complete. In the rest of this paper, we explain how we provide a full-KR approach to representing and performing the indexing KG with a declarative method allowing anyone to contribute to the definition and extraction of the indexed features.
%


 
\section{KRR Approach to Represent and Implement KG Indexing}

Several initiatives have integrated KR methods and tools into agile development methodologies to increase the adoption of these methods and tools. For instance, the ACIMOV methodology~\cite{hannou2023acimov} supports an agile-oriented and Git-based approach to ontology engineering.
In this paper, 
we extend this trend by proposing an approach based on IndeGx and on web-oriented forge services, such as GitHub, for collaborative and declarative definitions of rules and ontologies (pull requests for evolution, etc.), driving the indexing of KG endpoints in the wild. The collaboratively maintained rules guide a workflow periodically executed to update and publish the index as a dataset and service.
In this section, we cover the algorithms, heuristics, and optimizations that we designed to establish an operational index.

\subsection{Overview of the IndeGx framework}

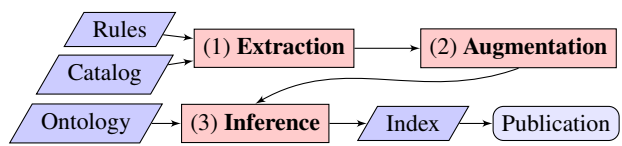
\begin{figure}[b]
    \centering
~\\
\begin{tikzpicture}
\node [data, fill=blue!20] at (0.2,1.25) (rules) {Rules};
\node [data, fill=blue!20] at (0,0.65) (seeds) {Catalog};
\node [process, fill=red!20] at (2.25,1) (extract) {(1) \textbf{Extraction}};
\node [process, fill=red!20] at (5.5,1) (augment) {(2) \textbf{Augmentation}};
\node [data, fill=blue!20] at (-0.25,0) (onto) {Ontology};
\node [process, fill=red!20] at (2,0) (infer) {(3) \textbf{Inference}};
\node [data, fill=blue!20] at (4.1,0) (index) {Index};
\node [terminator, fill=blue!10] at (6,0) (publish) {Publication};
\path [connector] (rules) -- (extract);
\path [connector] (seeds) -- (extract);
\path [connector] (extract) -- (augment);
\path [connector] (augment.south) to [out=200,in=30] (infer.north);
\path [connector] (onto) -- (infer);
\path [connector] (infer) -- (index);
\path [connector] (index) -- (publish);
\end{tikzpicture}
~\\
    \caption{Schema of the application of the IndeGx framework.}
    \label{fig:indegx_indexation}
\end{figure}

IndeGx takes as input a catalog of SPARQL endpoints and a set of rules. 
Then, the workflow consists of three steps presented in Figure~\ref{fig:indegx_indexation}.
IndeGx successively applies three processes to each endpoint to create a description. The whole process can be seen as a stratified reasoning approach where each step only depends on the previous ones.

\begin{enumerate}[label={(Step \arabic*)},leftmargin=1.25cm]
    \item The \textbf{extraction} rules extract information from the endpoint using SPARQL queries. These rules retrieve either existing content or compute information from the endpoint's content, as explained in Section~\ref{sec:extraction-rules}. 
    \item The \textbf{augmentation} step applies SPARQL rules to the KG resulting from step (1) to compute new information. As detailed in Section~\ref{sec:augmentation-rules}, these rules are applied locally, avoiding overloading remote SPARQL endpoints. 
    \item  An OWL RL reasoner performs \textbf{inferences} on the KG resulting from Step (3) using a given ontology. It mainly augments this KG with equivalent properties and inferred types as detailed in Section~\ref{sec:ontological-inferences}.
\end{enumerate}

This process results in an RDF KG indexing and describing all the SPARQL endpoints of KGs in the catalog. As explained in Section~\ref{sec:indegx-result}, the resulting KG also captures the traces of all the rules executed for its construction, making the whole process trackable and transparent.

On top of the IndeGx framework, 
KartoGraphI~\footnote{\url{http://prod-dekalog.inria.fr}}~\cite{maillot2022kartographi}, a Web application that exploits the KG produced by IndeGx to offer various visualizations, including the evolution of the indexed KGs through time, is provided.

Finally, to assist and encourage KG providers to include good-quality metadata in their knowledge graphs, 
there is an online service called Metadatamatic\footnote{\url{https://wimmics.github.io/voidmatic/}}~\cite{maillot2023metadatamatic}, that simplifies the creation of metadata for a dataset through a dynamic form and scripts that reuse the actions defined for IndeGx to extract statistics on the knowledge graphs.


\subsection{KR for Metadata Extraction in the Wild: Test \& Action Rules}
\label{sec:extraction-rules}
The extraction rules in IndeGx are inspired by the test suites, i.e., they comprise a series of \textit{tests} and \textit{actions} applied to each indexed KG and are represented in RDF.

\textit{Tests} are SPARQL queries sent to characterize the datasets and the endpoint of a KG. They often use the ASK clause to test the presence of a graph pattern in the KG. They are also used to test if an endpoint has the capacity for the application of more complex rules. Examples of \textit{tests} include checking the availability of an endpoint, checking the support of a given function (e.g., supports the {\tt VALUES} clause to bind a variable to a set of values), or confirming advertised statistics about the base (e.g., the number of triples). 

\textit{Actions} are SPARQL queries designed to generate a part of a KG's description. Typical \textit{actions} are SPARQL UPDATE queries designed to extract existing data from the KG's datasets, create new data according to the test result, or update an existing description. Examples of \textit{actions} include extracting the license asserted in a KG's dataset or computing the number of instances for each class.
In the IndeGx framework, the result of one \textit{test} can trigger zero, one, or more \textit{actions}. For instance, if an endpoint supports a specific SPARQL feature (e.g., the {\tt SERVICE} clause), several \textit{actions} based on that feature may be triggered (e.g., attempts to query known endpoints).




\begin{lstlisting}[language=rule, caption=Test and actions declaration for the extraction of the description of a dataset.,label=lst:dataset_description, style=codestyle]
<dataset_description.ttl> a kgi:TestQuery ,  mf:ManifestEntry ;
    dcterms:title "Presence of a Dataset." ;
    # Test query
    kgi:query """ASK {
        # $rawEndpointUrl is a keyword 
        # replaced by endpoint URL
        SERVICE $rawEndpointUrl {
            # Find a Dataset
            VALUES ?datasetClass {
                dcat:Dataset
                void:Dataset
            }
            ?kg a ?datasetClass .
        }
    }""" ;
    # Actions done if the test is positive
    kgi:onSuccess (
        [ mf:action """INSERT {
            ?kg a dcat:Dataset ;
                ?p ?o .
            } WHERE {
                SERVICE $rawEndpointUrl {
                    VALUES ?datasetClass {
                        dcat:Dataset
                        void:Dataset
                    }
                    ?kg a ?datasetClass ;
                        ?p ?o .
                }
            }"""
        ] ).
\end{lstlisting}

Listing~\ref{lst:dataset_description} is an example of a rule for extracting the description of datasets described in a KG. The rule is written in RDF as part of a test suite based on the \href{http://www.w3.org/2001/sw/DataAccess/tests/test-manifest#}{Manifest} and IndeGx vocabulary. The declaration of the test query starts on line 4. This test query checks whether the target KG contains at least one instance of a class \texttt{Dataset} (in DCAT or VOID vocabularies). If the result of the test query is {\tt true}, then the actions listed in the list starting on line 16 are applied. 
In this example rule, a single action applies a SPARQL UPDATE query, as on line 17.

Note that while Listing~\ref{lst:dataset_description} is an example of a rule for the extraction of the description of a dataset, it has been simplified to fit in this paper. In practice, several adaptations are needed for its application in the wild. Among others, the description may be stored in a named graph, making it necessary to search separately in the default and any named graph. The dataset instance may not be an IRI and thus may require generating a unique identifier to be stored in the index. Other optimizations include removing forbidden characters from IRIs or paginating the query for better performance.


These rules cover not only the extraction of characteristics of contents of KGs but also other characteristics such as the capabilities of their query engines and reasoners, their availability, etc. 
For instance, Listing~\ref{lst:not_exists} is a test declaration in which the capacity of the SPARQL endpoint of a KG is tested.
The result of this test does not depend on the result of the query declared on line 7. The query is not an ASK query as in Listing~\ref{lst:dataset_description}. In this case, the test is successful if the remote endpoint returns no errors. This specific test, inspired by SPARQLES~\cite{vandenbussche2017sparqles}, checks if a SPARQL endpoint will parse and accept a query containing the SPARQL 1.1 keywords {\tt NOT EXISTS}. The query in itself has been designed not to have any results as to have the shortest execution time possible if the server supports the keywords.


\begin{lstlisting}[language=rule, caption=Test and action declaration for the support of the NOT EXISTS clause.,label=lst:not_exists, style=codestyle]
@prefix ex: <http://example.org/> .
@prefix nonsense: <http://nonsense.com/> .
<SELNOTEXISTS.ttl> a kgi:TestQuery, mf:ManifestEntry ;
    dcterms:title "Support of NOT EXISTS." ;
    # Test query
    kgi:query """SELECT * {	
        SERVICE $rawEndpointUrl { 
            { SELECT * WHERE {
                ?s a ex:something .
                FILTER NOT EXISTS { ?s ex:property nonsense:1 } }
            } LIMIT 1
        }
    } """  ;
    # Action
    kgi:onSuccess (
        [ mf:action """INSERT DATA {
            $rawEndpointUrl sd:feature sparqles:SELNOTEXISTS .
            }""" 
        ] ) .
\end{lstlisting}


\subsection{SPARQL Rules for Data Augmentation}
\label{sec:augmentation-rules}
The augmentation rules are post-processing rules after the extraction and pre-processing rules before OWL reasoning.
These rules address the need to curate knowledge obtained on the open web and for specific data processing (e.g. numerical calculation or string manipulation).
SPARQL-based data augmentation rules add new knowledge to what was obtained using data extraction rules. They use the same SPARQL-based declaration as the data extraction rules, and they can also query an outside source that is different from the indexed KGs if needed. They complement the data extraction rules by transferring some computationally costly operations to the infrastructure of IndeGx's KG. They also perform transformations on the KG under construction that could not be covered by classical OWL reasoning (e.g. a numerical computation). 
Notably, data augmentation rules can estimate the cardinality of some sets which can then be used in OWL inferences, or correct low-level errors such as malformed URIs.



Listing~\ref{lst:equivalences} presents an example of a rule that captures a restricted form of equivalence only valid in the context of a dataset description and for properties used to declare a creation date. Contrary to Listings~\ref{lst:dataset_description} and \ref{lst:not_exists}, this rule is directly applied to IndeGx's KG without sending a query to a remote endpoint. As such, it does not need a test, and the {\tt kgi:DummyTest} on line 2 indicates that this rule's test is always positive. When describing a Dataset's creation date, this rule applies an equivalence between the four properties listed on lines 14 to 17. This restricted equivalence is outside the expressivity of OWL and thus pre-processed at that step. 

\begin{lstlisting}[language=rule, caption=Test and action for applying equivalences in the context of a dataset description.,label=lst:equivalences, style=codestyle]
<creation_date.ttl> a mf:ManifestEntry, kgi:DummyTest;
  # A "Dummy test" always succeed
  dcterms:title "Equivalences of creation dates properties for dataset." ;
  kgi:onSuccess (
    [ mf:action """INSERT {
          ?kg dcterms:created ?o ;
            schema:dateCreated ?o ;
            pav:createdOn ?o ;
            prov:generatedAtTime ?o .
        } WHERE {
          ?kg a dcat:Dataset ;
              ?p ?o .
          VALUES ?p { dcterms:created schema:dateCreated 
                      pav:createdOn prov:generatedAtTime }
        }""" 
    ] ) .
\end{lstlisting}


\subsection{OWL RL Ontology of the Index for Inferences}
\label{sec:ontological-inferences}

After the extraction and augmentation rules, OWL RL semantics rules are applied to the graph obtained from the previous steps and the IndeGx ontology.
This ontology provides a complementary way to enrich the knowledge graph with logical conclusions, this is useful when querying the index by declaring OWL-based augmentations.
The extraction and augmentation rules are challenging to read without good knowledge of SPARQL.
Therefore, complementary to these rules, the ontology contains all the augmentation rules that can be represented in OWL, e.g., defining concepts. 
The OWL definitions introduced in the ontology are typically useful to provide additional dimensions, making it easier to query and analyze the index's data.

\begin{lstlisting}[language=rule, caption=Declaration of a class that annotates endpoints that are SPARQL 1.1 compliant.,label=lst:sparql11, style=codestyle]
kgi:SPARQL11Compliant rdf:type owl:Class ;
  rdfs:label "SPARQL 1.1 compliant endpoint"@en ;
  rdfs:comment "indicates that an endpoint complies with SPARQL 1.1."@en ;
  rdfs:isDefinedBy kgi: ;
  rdfs:subClassOf sd:Service , dcat:DataService ;
  owl:equivalentClass [
    rdf:type owl:Restriction ;
    owl:onProperty sd:feature ;
    owl:hasValue sd:SPARQL11Query
  ] ,
  [ rdf:type owl:Class ;
    owl:intersectionOf (
      # List of tested SPARQL 1.1 features
      [ rdf:type owl:Restriction ;
        owl:onProperty sd:feature ;
        owl:hasValue sparqles:SELNOTEXISTS
      ]
      # ...
    )
  ] .
\end{lstlisting}

Listing~\ref{lst:sparql11} shows an example of a concept definition using restrictions in description logic to trigger a classification in the KG of the index. This definition will classify any KG's endpoint that has passed the tests, such as the one in Listing~\ref{lst:not_exists} that concerns features introduced by SPARQL1.1, on line 16. It will also annotate the endpoint with the property {\tt sd:feature sd:SPARQL11Query} on lines 8-9, defined by the SPARQL service description W3C recommendation.

\subsection{A KG of the Index and of the Indexing Process}
\label{sec:indegx-result}
The sequential application of the extraction rules, augmentation rules, and OWL rules incrementally enriches the data generated by IndeGx. This enables IndeGx to bypass third-party KG limitations, such as timeouts, and produce a high-quality KG containing detailed descriptions of KGs, including knowledge graph contents, query engine capabilities, entailment regimes, customizable quality metrics, non-functional properties (e.g., availability), etc.
The resulting graph also includes a declarative representation of the rules themselves that have been used to produce the index and the provenance information of all the generated KG descriptions.

Listing~\ref{lst:avg_execution_time} is an example of a rule reusing the declaration of other rules and the traces of the execution of IndeGx. The rule in this example is executed after the data augmentation rules. It aggregates the execution traces of the tests of each rule to obtain the average execution time of a SPARQL query by a given endpoint. The aggregation is performed by the query on line 6. The execution traces are accessible in IndeGx's KG in a dedicated named graph identified in line 8. The traces of the execution of a test are represented using the EARL and PROV-O ontologies between line 10 and line 15.

\begin{lstlisting}[language=rule,caption=Test and action declaration for a rule using the trace of the execution of queries to compute the average execution time on a SPARQL endpoint., label=lst:avg_execution_time, style=codestyle]
<execution_time_average.ttl> a mf:manifestEntry , kgi:DummyTest ;
    kgi:onSuccess (
    [ mf:action """INSERT {
        ?endpointUrl kgi:averageExecutionTime ?averageDuration .
    } WHERE {
        { SELECT ?endpointUrl (AVG(?duration) AS ?averageDuration) {
          # Execution traces are stored in a dedicated graph
          GRAPH kgi:Logs {
            kgi:Logs kgi:trace ?testTrace .
            ?testTrace a prov:Activity ;
              earl:result ?testResult ;
              earl:subject ?endpointUrl ;
              prov:startedAtTime ?startTime ;
              prov:endedAtTime ?endTime .
            ?testResult earl:outcome earl:passed .
          }
        # Computation of the duration of the query execution
        BIND((?endTime - ?startTime) AS ?duration)
        } GROUP BY ?endpointUrl }
    }"""
    ] ) .
\end{lstlisting}
%

 
\section{Experiment: Indexing World ``Wild'' SPARQL Endpoints}
\label{sec:experiment}

The entire framework with all the rules and knowledge graphs described in the previous section has been implemented and published online\footnote{\url{https://github.com/Wimmics/IndeGx/}} by the authors of \cite{maillot2023indegx} to provide a fully functional public index of KGs available on the web and their characteristics.
The resulting KG can be used to identify sources for any application that consumes RDF graphs. It can also be used to provide benchmarks and testbeds and is helpful for KR researchers and domain experts 
to understand KR practices in the real world ``wild'' web, providing statistics on the KR primitives, ontologies, and formalisms used in each KG.
In the next sections, we illustrate some insights we gained by analyzing the index.
Please note that each experiment was done at a different time and that some endpoints may not have been available for all experiments. 
The reality of KR in the wild is that, by combining several methods and sources, we were able to identify 870 SPARQL endpoints out of which only an average of 300 were available at a given time. Only the endpoints active at the moment of an experiment are considered for that experiment. 
\subsection{The Catalog of Knowledge Sources}
\label{sec:experiment_catalog}
The index starts from a catalog of endpoints aggregated from multiple sources, such as Wikidata, the LOD Cloud, search campaigns on search engines, and other monitoring approaches. 
The 
catalog available online\footnote{As the catalog contains authorship metadata, the link has been redacted for the review process}
contains 870 endpoint URLs, of which, on average, 300 were responsive during each experimentation. 
Each campaign attempts to obtain a maximum amount of metadata about these endpoints.

\begin{figure}[ht]
    \centering
    \frame{\includegraphics[width=0.98\linewidth]{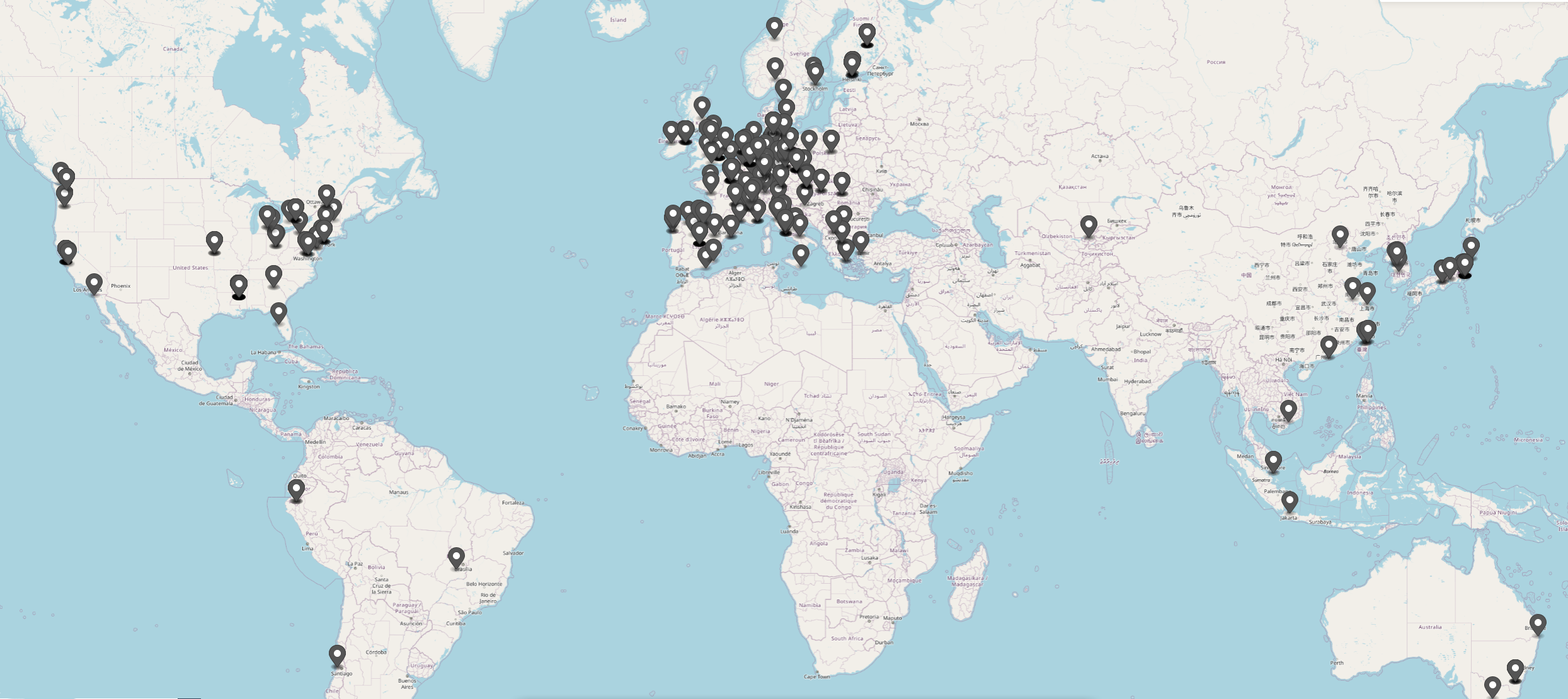}}
    \caption{Map of the geolocations of the tested endpoints.}
    \label{fig:geolocation_map}
\end{figure}

Using external tools, we retrieved the GPS coordinates of the endpoints' hosting servers as an example of metadata characterizing the catalog.
Figure~\ref{fig:geolocation_map} shows the geolocation of the endpoints, and we can see that the endpoints are hosted mainly in Europe, North America, and East Asia. This distribution can be explained by multiple factors, such as the adoption of Semantic Web technologies being pushed unequally worldwide and/or a bias in the sources used to build the catalog.
Among other metadata of interest for an endpoint, we can also mention the links between KGs, application domains, etc.

\subsection{Content of Knowledge Sources}
\label{sec:experiment_content}
\begin{figure}[ht]
    \centering
    \includegraphics[width=.9\linewidth]{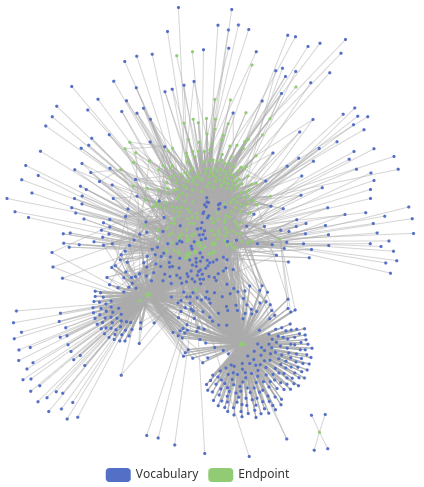}
    \caption{Graph of the links between endpoints and the vocabularies they use.} 
    \label{fig:endpoint_vocabularies_graph}
\end{figure}

One way to characterize the content of a knowledge graph is to identify the vocabularies it uses (e.g. RDFS / OWL ontologies, SKOS thesauri). 
We use the vocabulary descriptions available in the Linked Open Vocabularies (LOV) Endpoint~\cite{vandenbussche2017LOV} and the prefix-namespace registry service  \href{https://prefix.cc}{Prefix.cc} to determine the number of known vocabularies used in each endpoint. Figure~\ref{fig:endpoint_vocabularies_graph} shows the graph of endpoints in green and vocabularies in blue in a force layout. The edges indicate that at least one vocabulary element is used in a KG. In the figure, widely used meta-ontologies (such as RDFS and OWL metamodels) and very popular ontologies (such as Schema.org) appear in the middle of a cluster of endpoints. Sparsely used vocabularies appear outside the group, as they are only used in one or two endpoints. The two endpoints that are repositories of vocabularies, namely the mirrors of LOV and the Openlink Uriburner KG, appear surrounded by vocabularies (see the two green nodes surrounded by blue nodes at the bottom of the figure).

Among other indicators of a KG's content, we also have the presence of links to other datasets, the languages used in literals, the size of its datasets in terms of triples, resources, etc. 

\subsection{Provenance of Knowledge}
\label{sec:experiment_prov}
One key characteristic in deciding to (re)use a knowledge source is its provenance.
We extract the provenance information available in a KG, taking into account the fact that this information can be described using either the vocabularies recommended by the W3C, such as VoID and DCAT, or a general ontology, such as Schema.org. 
We designed rules to extract the most frequent ways of declaring the provenance of KGs.
Tables~\ref{tab:dcat-void-classes} and~\ref{tab:dcat-void-properties} show the ontological primitives we found in the descriptions of KG's endpoints and their frequency. We can notice, for instance, that for the 106 instances of {\tt void:Dataset}, the access URL of knowledge graphs is given with 3 different properties, i.e. {\tt dcat:downloadURL}, {\tt dcat:accessURL} and {\tt void:dataDump}, in relatively the same number of KGs. Conversely, the URL of endpoints are pejoratively given using the property {\tt void:sparqlEndpoint} while the {\tt dcat:endpointURL} is only used 26 times. We see that spatial-related information such as {\tt centroid} or {\tt dcat:bbox} and time-related information, such as {\tt dcat:endDate}, are rarely present. Generally, we can expect to find general information in the description of a KG, such as the vocabulary used or the keywords describing the dataset. Still, we can rarely find more specific details. We also see that both DCAT and VoID should be looked for when searching for the description of KGs, along with other more general vocabularies such as DCTerms and PROV-O.

\begin{figure}[ht]
    \centering
    \includegraphics[width=0.95\linewidth]{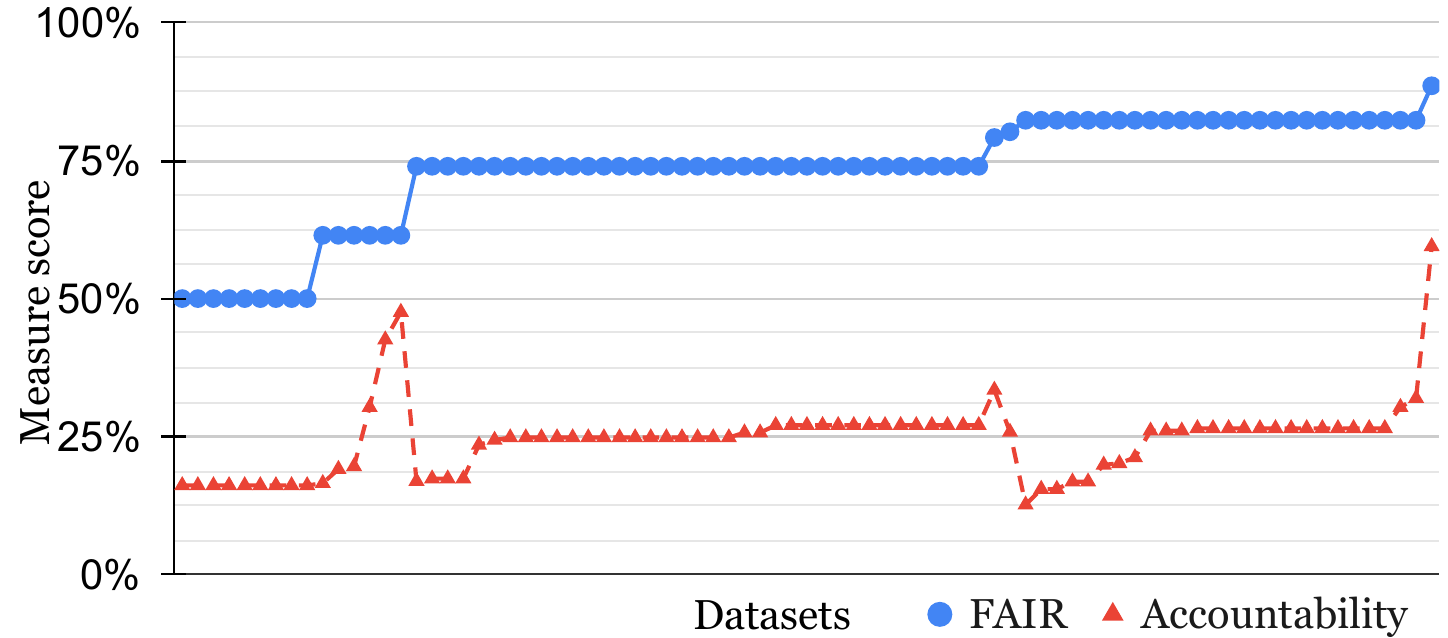}
    \caption{FAIRness and accountability scores in ascending FAIRness score order. } 
    \label{fig:FAIR_Acc_comparison}
\end{figure}

We evaluated the descriptions that were extracted according to two dataset quality measures
: FAIRness~\cite{gaignard2023fair} and accountability~\cite{andersen2023framework}. 
They are different takes on the evaluation of the quality and transparency of datasets. FAIRness, in the FAIR-Checker interpretation, evaluates general metrics such as LOD good practices (e.g. dereferenceability of URIs, minimum number of known vocabularies) 
or its registration on known websites. 
By contrast, the accountability checks the conjunction of specific pieces of information, "whatever" the vocabulary.
The two measures were implemented as rules in the IndeGx framework and evaluated over 339 endpoints. The results showed that in 25 endpoints, we could find 80 datasets with enough provenance metadata to compare the two measures. Figure~\ref{fig:FAIR_Acc_comparison} shows the evaluation results in ascending FAIRness score order.

\begin{table}[ht!]
\caption{Number of endpoints using each DCAT \& VoID class}
\center \small
\begin{tabular}{|l||r|}
\hline
\textbf{Class}        & \textbf{\# endpoints} \\ \hline \hline
void:Dataset  & 106  \\ \hline 
dcat:Dataset& 80   \\ \hline
dcat:Distribution     & 65   \\ \hline
dcat:Catalog& 46   \\ \hline
void:Linkset  & 39   \\ \hline
dcat:DataService      & 25   \\ \hline
void:DatasetDescription & 19   \\ \hline
dcat:Resource         & 18   \\ \hline
dcat:CatalogRecord    & 14   \\ \hline
void:TechnicalFeature   & 10   \\ \hline
dcat:Role   & 6    \\ \hline
dcat:Relationship     & 3    \\ \hline
\end{tabular}
\label{tab:dcat-void-classes}
\end{table}

\begin{table}[ht!]
\caption{Number of endpoints using each DCAT \& VoID property}
\begin{minipage}{0.5\textwidth}
\center 
\small
\begin{tabular}{|l|r|}
\hline
\textbf{Property}        & \textbf{\# endpoints} \\ \hline \hline
void:sparqlEndpoint & 79   \\ \hline
void:vocabulary     & 65   \\ \hline
dcat:keyword      & 56   \\ \hline
dcat:distribution & 56   \\ \hline
void:triples        & 56   \\ \hline
dcat:downloadURL  & 50   \\ \hline
dcat:accessURL    & 49   \\ \hline
void:dataDump       & 47   \\ \hline
dcat:mediaType    & 45   \\ \hline
void:subset         & 45   \\ \hline
dcat:theme        & 44   \\ \hline
dcat:landingPage  & 42 \\ \hline
void:exampleResource& 42   \\ \hline
dcat:dataset     & 40   \\ \hline
dcat:contactPoint& 39   \\ \hline
void:uriSpace      & 39   \\ \hline
void:inDataset     & 38   \\ \hline
void:entities      & 36   \\ \hline
void:linkPredicate & 33   \\ \hline
void:classes       & 32   \\ \hline
void:distinctSubjects         & 28   \\ \hline
void:properties    & 27   \\ \hline
void:distinctObjects& 27   \\ \hline
dcat:endpointURL & 26   \\ \hline
void:feature       & 26   \\ \hline
void:class         & 26   \\ \hline
void:subjectsTarget& 25   \\ \hline
void:objectsTarget & 25   \\ \hline
\end{tabular}
\end{minipage}
\begin{minipage}{0.5\textwidth}
\center 
\small
\begin{tabular}{|l|r|}
\hline
\textbf{Property}        & \textbf{\# endpoints} \\ \hline \hline
void:classPartition& 25   \\ \hline
void:propertyPartition        & 24   \\ \hline
void:property      & 24   \\ \hline
dcat:servesDataset& 20   \\ \hline
void:uriRegexPattern& 20   \\ \hline
dcat:byteSize    & 18   \\ \hline
void:rootResource  & 18   \\ \hline
dcat:themeTaxonomy& 15   \\ \hline
void:target        & 15   \\ \hline
dcat:accessService& 13   \\ \hline
dcat:temporalResolution     & 12   \\ \hline
dcat:startDate   & 12   \\ \hline
dcat:record      & 11   \\ \hline
dcat:endpointDescription    & 11   \\ \hline
dcat:compressFormat         & 11   \\ \hline
void:uriLookupEndpoint        & 11   \\ \hline
dcat:endDate     & 9    \\ \hline
dcat:service     & 7    \\ \hline
dcat:packageFormat& 5    \\ \hline
dcat:spatialResolutionInMeters        & 4    \\ \hline
dcat:catalog     & 4    \\ \hline
dcat:bbox        & 4    \\ \hline
void:openSearchDescription    & 4    \\ \hline
dcat:qualifiedRelation      & 3    \\ \hline
dcat:hadRole     & 3    \\ \hline
dcat:centroid    & 3    \\ \hline
void:documents     & 1   \\  \hline
\end{tabular}
\end{minipage}
\label{tab:dcat-void-properties}
\end{table}

\subsection{Languages and Fragments Used}
\label{sec:experiment_vocabs}
From a KR perspective, it is interesting to know which language, or which profile or fragment of a language, is effectively used in the wild to formalize and publish knowledge graphs. This shows the adoption and the required reasoning capabilities to exploit each source. We tested the presence of the classes and properties of the most well-known ``meta-vocabulary'' among 320 endpoints. Each class and property presence was tested with an efficient ASK query.
The 320 endpoints tested were taken from the initial catalog and were available during the experiment in April 2024. 

\begin{figure}[ht]
    \centering
    \includegraphics[width=.95\linewidth]{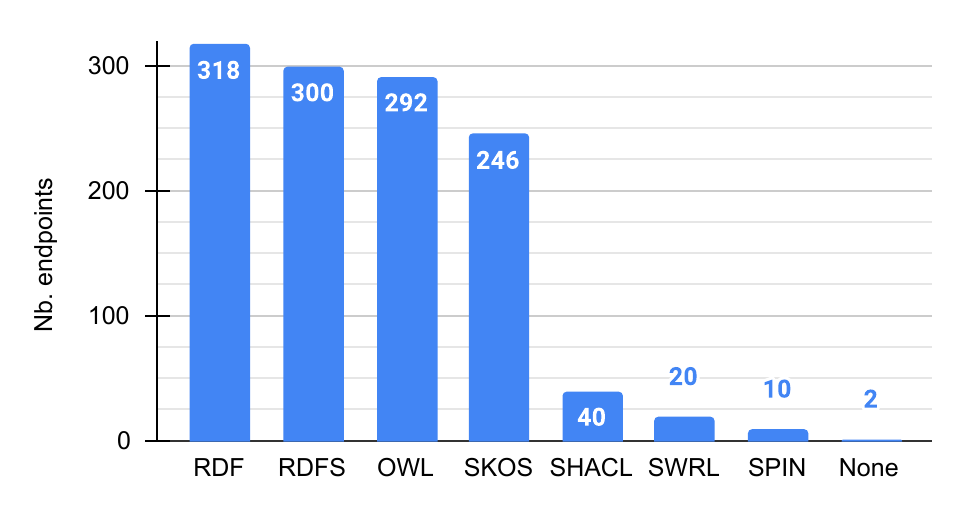}
    \caption{Number of endpoints using each meta-vocabulary}
    \label{fig:metavocabularies}
\end{figure}

For each meta-vocabulary, 
Figure~\ref{fig:metavocabularies} shows the endpoints containing at least one of their primitives. RDF, RDFS, OWL, and SKOS are well ahead regarding usage. We also see that two endpoints were using none of those meta-vocabularies: one of them only answered the availability test query, and the other used exclusively classes and properties from Schema.org.
We indirectly tested the application of basic RDFS entailment among 317 active endpoints. As any resource in a KG with RDFS entailment should be an instance of {\tt rdfs:Resource}, we tested the presence of at least one instance of {\tt rdfs:Resource}. In 265 endpoints, there was no such triple. This shows that at least 83\% of publicly available endpoints do not support basic RDFS entailment.

\subsection{Knowledge Quality Features}
\label{sec:experiment_quality}
To decide the knowledge sources we want to rely on, we often have several criteria for quality and general characteristics that we would like them to meet. In the literature, several criteria have been identified as indicators of the quality of a KG, such as those presented in the survey~\cite{xue2022KnowledgeGraphQuality}. IndeGx proposes a first set of rules to evaluate KGs according to several existing quality criteria and supports the customization and extensions of this rule set at will. 

For instance, best practices for RDF data structuring can be measured using SPARQL queries. They give clear quality indicators for given use cases, such as the proportion of resources with a human-readable label for readability or the proportion of string literals with a language tag. As shown in 
related works
\cite{maillot2023indegx,maillot2022kartographi}, an average of 42\% of the resources have no label. The nature of some of those quality indicators depends on the community. For example, using blank nodes and RDF data structures such as lists, sequences, and bags poses problems for some data aggregation methods. However, OWL and SHACL use blank nodes and data structures to define vocabularies. For this reason, the relevance of that indicator depends on the targeted task and the nature of the considered graph.

More generally, the features presented in other sections, such as the presence of provenance information in section~\ref{sec:experiment_prov} or the usage of specific vocabularies in section~\ref{sec:experiment_vocabs}, are also used as quality measures for a KG in some communities. As the definition of quality depends on the usage of the KG, many characteristics of a KG could be used as a quality factor.

\subsection{Knowledge Graph System Characteristics}
\label{sec:experiment_system}
In addition to characterizing the knowledge graphs themselves, the capabilities and limits of the underlying infrastructure, software, and implementation may also be important before establishing a dependency on a knowledge source. For this reason, IndeGx allows us to study not only the content of KGs but also the capacities of their endpoints.

SPARQL queries enable us to test each of the SPARQL features that SPARQL endpoints may support in a manner inspired by SPARQLES~\cite{vandenbussche2017sparqles}. Completion rules can test each feature and annotate the endpoint accordingly. As shown in listing~\ref{lst:sparql11}, an inference may annotate each endpoint according to the conjunction of features that they support. The results of these tests are published in 
(redacted for peer review)
, showing, among others, that more than 60\% of the endpoints support most of SPARQL 1.1, and that the SERVICE keyword used for federated queries is the least supported SPARQL feature.

The IndeGx framework also tracks each endpoint's time to answer each query. This enables the computation of a general response time for endpoints, depending on the type of queries sent. Furthermore, repeated indexing through time enables monitoring an endpoint's availability.

In April 2024, we tested the behavior of 317 active SPARQL endpoints to generate two random numbers and of the current date twice in the same query. According to the SPARQL recommendation, two calls to the {\tt RAND} function in the same SPARQL query should return two different results, and two calls to the {\tt NOW} function in the same query should return the same date. The results show that 49\% of the endpoints will return the same number when asked for two different random number generations in the same query, and 6\% of the endpoints will return two different dates when asked for the current date twice in the same query.

\begin{figure}[ht!]
    \centering
    \includegraphics[width=.95\linewidth]{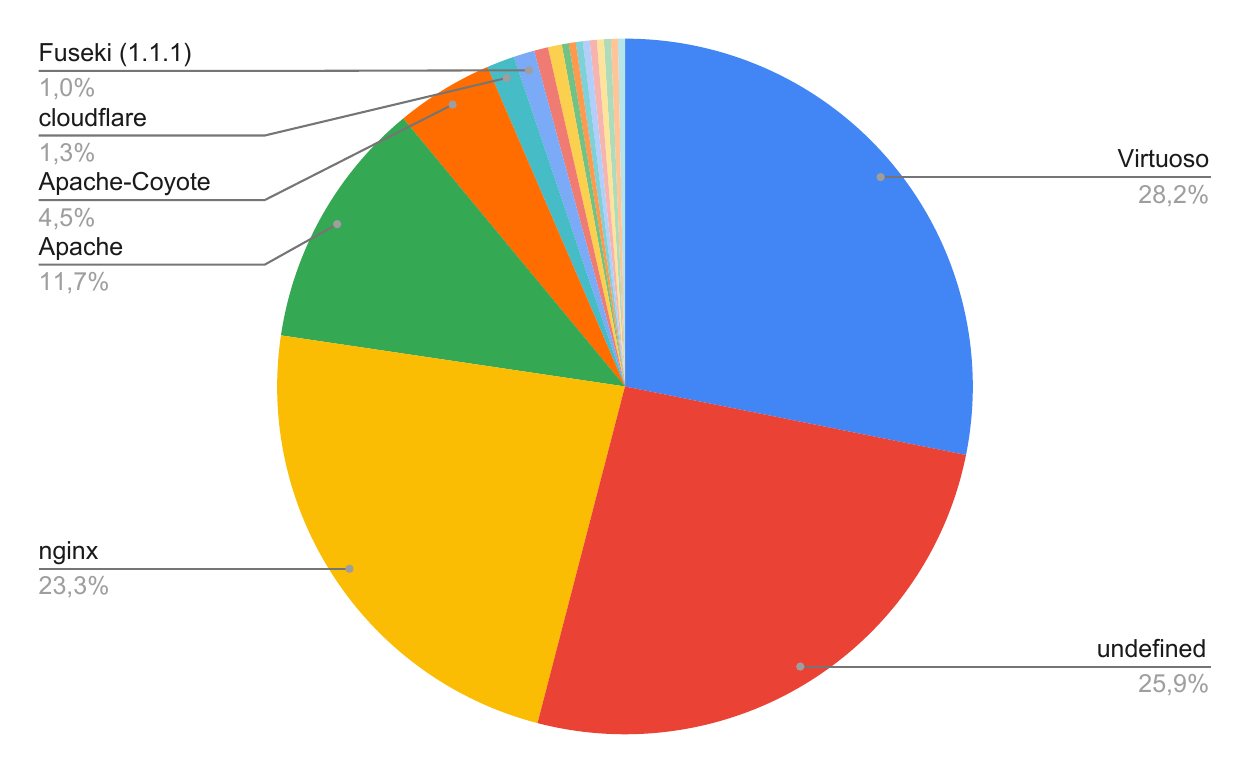}
    \caption{Server HTTP header values for SPARQL endpoints on April 2024.}
    \label{fig:http_header}
\end{figure}

Using a non-standard extension of SPARQL, LDScript~\cite{corby2021service}, we also obtained the type of server declared by the SPARQL endpoint in HTTP headers. 
Figure~\ref{fig:http_header} shows the distribution of the \texttt{server} HTTP header values. Apart from 28.2\% of Virtuoso servers and 1\% of Apache Fuseki servers, most SPARQL endpoints do not return the kind of SPARQL engines they use, or a front-end web server removes the header. In the detailed values of the headers, there are 31 different versions of Virtuoso servers, 18 versions of Nginx, and 7 Apache server versions. 

The differences between endpoints in SPARQL coverage and server version paint a picture of a great diversity of behavior in the KGs available through SPARQL endpoints.
Using LDScript also enables IndeGx to detect the usage of dereferenceable URIs in datasets, among other features.

 
\section{Conclusion}

In this article, we showed that we can successfully deploy KR formalisms and reasoning means to create an index of online knowledge graphs. 
We contributed a new open-source KR tool 
based on the IndeGx framework and a new resource, as the index is a publicly available KG for identifying and accessing all the indexed datasets.

This KR extension of the IndeGx framework enabled new ways to annotate the data retrieved from KGs and enhanced the framework's usability in general.

We also conducted multiple experiments over publicly available KGs with SPARQL endpoints. Each experiment was done with a different use case to demonstrate versatility and provide new insights about the KGs' contents, statuses, and endpoint capabilities. The results of our experiments show a great diversity in the world wild semantic web, in which the type of content, the vocabulary used, the endpoint capabilities, and the general quality of KG can differ significantly. These results can also help us identify KR research opportunities, resources, and challenges.

The metadata retrieved by our approach, and the new insights they brought, constitute critical resources for current active area of research, such as source selection for federated queries or GraphRAG approaches, neuro-symbolic AI approaches requiring specific ontological constructs, and in general for a analysing the state of the data and services available (including, e.g., geographical biases, lack of technological updates). For example, knowing the main triplestores and the specific choices made in their query engine implementation allows us to explain some results obtained in the wild (e.g., different behaviors of the RAND and NOW functions, etc.) and anticipate the impact on any application calling them.

\section{Acknowledgments}
This work is supported by the DeKaloG project, ANR-19-CE23-0014, and the 3IA Côte d'Azur ANR-19-P3IA-0002.

\bibliographystyle{kr}
\bibliography{citations}

\begin{thebibliography}{}

\bibitem[\protect\citeauthoryear{Andersen \bgroup et al\mbox.\egroup
  }{2023}]{andersen2023framework}
Andersen, J.; Cazalens, S.; Lamarre, P.; and Maillot, P.
\newblock 2023.
\newblock A framework to assess knowledge graphs accountability.
\newblock In {\em 2023 IEEE International Conference on Web Intelligence and
  Intelligent Agent Technology (WI-IAT)},  213--220.
\newblock IEEE.

\bibitem[\protect\citeauthoryear{Brinkmann, Primpeli, and
  Bizer}{2023}]{brinkmann2023webdatacommonsschema}
Brinkmann, A.; Primpeli, A.; and Bizer, C.
\newblock 2023.
\newblock {The Web Data Commons Schema.org Data Set Series}.
\newblock In {\em Companion Proceedings of the ACM Web Conference 2023}, WWW
  '23 Companion,  136–139.
\newblock New York, NY, USA: Association for Computing Machinery.

\bibitem[\protect\citeauthoryear{Corby \bgroup et al\mbox.\egroup
  }{2021}]{corby2021service}
Corby, O.; Faron, C.; Gandon, F.; Graux, D.; and Michel, F.
\newblock 2021.
\newblock {Beyond Classical {SERVICE} Clause in Federated {SPARQL} Queries:
  Leveraging the Full Potential of {URI} Parameters}.
\newblock In Mayo, F. J.~D.; Marchiori, M.; and Filipe, J., eds., {\em
  Proceedings of the 17th International Conference on Web Information Systems
  and Technologies, {WEBIST} 2021, October 26-28, 2021},  65--76.
\newblock {SCITEPRESS}.

\bibitem[\protect\citeauthoryear{Gaignard \bgroup et al\mbox.\egroup
  }{2023}]{gaignard2023fair}
Gaignard, A.; Rosnet, T.; De~Lamotte, F.; Lefort, V.; and Devignes, M.-D.
\newblock 2023.
\newblock {FAIR-Checker: supporting digital resource findability and reuse with
  Knowledge Graphs and Semantic Web standards}.
\newblock {\em Journal of Biomedical Semantics} 14(1):1--14.

\bibitem[\protect\citeauthoryear{{Glimm} \bgroup et al\mbox.\egroup
  }{2012}]{glimm2012OWLArriveWeb}
{Glimm}, B.; {Hogan}, A.; {Kr{\"o}tzsch}, M.; and {Polleres}, A.
\newblock 2012.
\newblock {OWL: Yet to arrive on the Web of Data?}
\newblock {\em arXiv e-prints}  arXiv:1202.0984.

\bibitem[\protect\citeauthoryear{Hannou \bgroup et al\mbox.\egroup
  }{2023}]{hannou2023acimov}
Hannou, F.-Z.; Charpenay, V.; Lefran{\c{c}}ois, M.; Roussey, C.; Zimmermann,
  A.; and Gandon, F.
\newblock 2023.
\newblock {The ACIMOV Methodology: Agile and Continuous Integration for Modular
  Ontologies and Vocabularies}.
\newblock In {\em MK 2023-2nd Workshop on Modular Knowledge associated with
  FOIS 2023-the 13th International Conference on Formal Ontology in Information
  Systems}.

\bibitem[\protect\citeauthoryear{Kopsachilis \bgroup et al\mbox.\egroup
  }{2020}]{kopsachilis2020geolod}
Kopsachilis, V.; Vaitis, M.; Mamoulis, N.; and Kotzinos, D.
\newblock 2020.
\newblock Recommending {{Geo-semantically Related Classes}} for {{Link
  Discovery}}.
\newblock {\em Journal on Data Semantics} 9(4):151--177.

\bibitem[\protect\citeauthoryear{Maillot \bgroup et al\mbox.\egroup
  }{2022}]{maillot2022kartographi}
Maillot, P.; Corby, O.; Faron, C.; Gandon, F.; and Michel, F.
\newblock 2022.
\newblock {KartoGraphI: Drawing a Map of Linked Data}.
\newblock In {\em Extended Semantic Web Conference}.
\newblock Springer.

\bibitem[\protect\citeauthoryear{Maillot \bgroup et al\mbox.\egroup
  }{2023a}]{maillot2023indegx}
Maillot, P.; Corby, O.; Faron, C.; Gandon, F.; and Michel, F.
\newblock 2023a.
\newblock {IndeGx: A model and a framework for indexing RDF knowledge graphs
  with SPARQL-based test suits}.
\newblock {\em Journal of Web Semantics} 76:100775.

\bibitem[\protect\citeauthoryear{Maillot \bgroup et al\mbox.\egroup
  }{2023b}]{maillot2023metadatamatic}
Maillot, P.; Corby, O.; Faron, C.; Gandon, F.; and Michel, F.
\newblock 2023b.
\newblock {Metadatamatic: A Web application to Create a Dataset Description}.
\newblock In {\em Companion Proceedings of the ACM Web Conference 2023},
  123--126.

\bibitem[\protect\citeauthoryear{Polleres \bgroup et al\mbox.\egroup
  }{2020}]{polleres2020MoreDecentralizedVision}
Polleres, A.; Kamdar, M.~R.; Fern{\'a}ndez, J.~D.; Tudorache, T.; and Musen,
  M.~A.
\newblock 2020.
\newblock A more decentralized vision for {{Linked Data}}.
\newblock {\em Semantic Web} 11(1):101--113.

\bibitem[\protect\citeauthoryear{Shi \bgroup et al\mbox.\egroup
  }{2023}]{shi2023VOYAGELargeCollection}
Shi, Q.; Wang, J.; Pan, J.~Z.; and Cheng, G.
\newblock 2023.
\newblock {{VOYAGE}}: {{A Large Collection}} of~{{Vocabulary Usage}} in~{{Open
  RDF Datasets}}.
\newblock In Payne, T.~R.; Presutti, V.; Qi, G.; {Poveda-Villal{\'o}n}, M.;
  Stoilos, G.; Hollink, L.; Kaoudi, Z.; Cheng, G.; and Li, J., eds., {\em The
  {{Semantic Web}} -- {{ISWC}} 2023},  211--229.
\newblock Cham: Springer Nature Switzerland.

\bibitem[\protect\citeauthoryear{Vandenbussche \bgroup et al\mbox.\egroup
  }{2017a}]{vandenbussche2017LOV}
Vandenbussche, P.-Y.; Atemezing, G.~A.; Poveda-Villal{\'o}n, M.; and Vatant, B.
\newblock 2017a.
\newblock {Linked Open Vocabularies (LOV): a gateway to reusable semantic
  vocabularies on the Web}.
\newblock {\em Semantic Web} 8(3):437--452.

\bibitem[\protect\citeauthoryear{Vandenbussche \bgroup et al\mbox.\egroup
  }{2017b}]{vandenbussche2017sparqles}
Vandenbussche, P.-Y.; Umbrich, J.; Matteis, L.; Hogan, A.; and Buil-Aranda, C.
\newblock 2017b.
\newblock {SPARQLES: Monitoring public SPARQL endpoints}.
\newblock {\em Semantic web} 8(6):1049--1065.

\bibitem[\protect\citeauthoryear{Xue and
  Zou}{2022}]{xue2022KnowledgeGraphQuality}
Xue, B., and Zou, L.
\newblock 2022.
\newblock Knowledge {{Graph Quality Management}}: A {{Comprehensive Survey}}.
\newblock {\em IEEE Transactions on Knowledge and Data Engineering}  1--1.

\bibitem[\protect\citeauthoryear{Yamamoto, Yamaguchi, and
  Splendiani}{2018}]{yamamoto2018yummydata}
Yamamoto, Y.; Yamaguchi, A.; and Splendiani, A.
\newblock 2018.
\newblock {YummyData: providing high-quality open life science data}.
\newblock {\em Database} 2018:bay022.

\end{thebibliography}

\end{document}